# Quantifying Prosodic Variability in Middle English Alliterative Poetry

Roger Bilisoly[1]

[1]Department of Mathematical Sciences, Central Connecticut State University, 1615 Stanley St, New Britain, CT 06050-4010

**Abstract**
Interest in the mathematical structure of poetry dates back to at least the 19[th] century: after retiring from his mathematics position, J. J. Sylvester wrote a book on prosody called *The Laws of Verse*. Today there is interest in the computer analysis of poems, and this paper discusses how a statistical approach can be applied to this task. Starting with the definition of what Middle English alliteration is, *Sir Gawain and the Green Knight* and William Langland's *Piers Plowman* are used to illustrate the methodology. Theory first developed for analyzing data from a Riemannian manifold turns out to be applicable to strings allowing one to compute a generalized mean and variance for textual data, which is applied to the poems above. The ratio of these two variances produces the analogue of the F test, and resampling allows p-values to be estimated. Consequently, this methodology provides a way to compare prosodic variability between two texts.

## 1. Introduction

The ultimate goal of this paper is to quantify the variability of stress patterns of Middle English alliterative poems. As a concrete example, William Langland's *Piers Plowman* and *Sir Gawain and the Green Knight*, the author of which is unknown, are compared. This is done in two ways. First, the words of the initial 220 alliterating Middle English lines of each poem are labeled as either alliterating (coded as 1) or not (coded as 0), so each line can be viewed as a string (or vector) of 0s and 1s. Finally, a technique is developed to compute the variability of these binary strings. Second, prosodic structure for alliterative poetry is traditionally represented by strings such as "aa/ax" or "ab/ab." Here the forward slash stands for a caesura, a pause near the middle of the line. Letters stand for stressed words, where *a* and *b* each represent alliterating words with distinct initial sounds, while *x* stands for a non-alliterating stressed word. It turns out that the above technique for binary strings is applicable to this notation, too.

In older literary studies the unstressed words were often ignored, which is the reason why prosodic coding such as "aa/ax" only refers to the stressed ones. Chapter 1 of Cole (2007) gives an overview of this earlier work, and her dissertation goes on to show that the lines of poems such as *Piers Plowman* and *Sir Gawain* do exhibit meter, which inspired the author to analyze binary coding vectors that represent all words in a line in this paper.

## 1.1 The Structure of Middle English Alliterative Poetry

The metric structure of Middle English alliterative poetry is complex and there have been many debates on what rules exist and how they are applied. Cole (2007) gives an overview of earlier work by researchers such as Thomas Cable and Hoyt Duggan. However, her dissertation argues that all the syllables in a line should be considered, and she uses the traditional prosodic notation of strings composed of /s and *x*s, which represent stressed and unstressed syllables, respectively. However, this paper focuses on meter at the level of words.

Figure 1 shows the first ten lines of the prologue of *Piers Plowman*, with the alliterating words in bold font (added by the author), and the forward slash represents a caesura, which splits each line into two pieces: the a-verse and the b-verse. Although the full set of rules is debated and not completely known, for this paper the following simplified rules are sufficient. First, only stressed content words are eligible for alliteration. So function words such as articles, prepositions, pronouns, and so forth, are excluded. Second, alliteration means that initial consonant sounds are repeated. Although vowels can alliterate, this is relatively uncommon, and note that phonology is key, not the spelling. Third, unstressed initial syllables are ignored. For example, in line 6, "bifel" alliterates with the other f-initial words because the prefix "bi" is not stressed. Fourth, there are some special cases. For example, the sh-sound in line 2 is considered a separate case from /s/.

> In a **somer seson** / whan **softe** was the **sonne**,
> I **shoop** me into **shroudes /** as I a **sheep** were,
> In **habite** as an **heremite** / **unholy** of werkes,
> Wente **wide** in this **world** / **wondres** to here.          4
> Ac on a **May morwenynge /** on **Malverne** hilles
> Me **bifel** a **ferly** / of **Fairye** me thoghte.
> I was **wery forwandred /** and **wente** me to reste
> Under a **brood bank /** by a **bourne** syde;                   8
> And as I **lay** and **lenede /** and **loked** on the watres,
> I **slombred** into a **slepyng** / it **sweyed** so murye.

**Figure 1:** These are the first ten lines of the prologue of *Piers Plowman* from Langland et al. (1978). The stressed and alliterating words were bolded by the author, and the slashes are caesura, which are taken from Langland and Skeat (1867-85).

For the modern reader, Middle English alliterative poetry is unfamiliar, both in style and in spelling. However, there is one example in modern English by J. R. R. Tolkien (*The Fall of Arthur*, Tolkien (2013)), who is best known for *The Hobbit* and *The Lord of the Rings*, but who was a professor of Anglo-Saxon and wrote about and edited Old English and Middle English alliterative poetry (for example, Anonymous, Tolkien, and Gordon (1967)).

## 1.2 Middle English Phonology

As noted above, alliteration is based on the initial sound of a word, not its spelling. For example, "cat," "king," and "quit" all alliterate, but "cat," "cent," and "ciao" do not. Consequently, the phonology of Middle English is needed, which cannot be known with certainty, but historical linguists are in general agreement on many issues. Moreover, only the correspondences between initial spellings and initial sounds are required here, and because of phonotactic constraints, there are not that many possibilities.

For this study, the following rules are sufficient, which are simplified versions of those given in Chapter 4 of Upward and Davidson (2011) and Chapter 8 of Brinton and Arnovick (2006). First, as is true in modern English, the initial letter *c* has a hard and soft version. Hard-c, which has a /k/ sound, happens before back vowels and consonants, and soft-c (an /s/ sound) before the rest of the vowels. Second, the initial-g also has a hard and soft version, as does its predecessor, *ȝ* (the Middle English letter yogh, which died out under the influence of Norman scribes), with roughly the same rules as the letter *c*. Third, Middle English has more initial consonant clusters than today, and none of the letters are silent. For example, words with initial-kn have both sounds pronounced.

## 2. Coding Prosodic Structure

In the study of poetic prosody, although specific terms are used, good poets often purposely deviate from strict adherence to a given meter. For example, iambic pentameter is supposedly composed of ten syllables per line where the stress is on the even numbered ones, usually coded as "x/x/x/x/x/." However, when one looks at examples, this is often not the case. For instance, Hamlet's famous line "To be, or not to be, that is the question," has eleven syllables, so the rules are more complex than the term *iambic pentameter* suggests. As the title of Section 1.3 of Fabb (2002) states, "there are ten metrified syllables in the iambic pentameter line," which is true for what he calls projected syllables, so "x/x/x/x/x/" does hold, but for only a certain type of syllable.

There is, however, much more variability in the extant Middle English alliterative texts. This arose in many ways: lack of standardization of Middle English itself; the use of scribes to copy texts before the age of printing in England started by William Caxton; the variety of dialects that existed then, and so forth. Even in the two simplifying coding schemes defined below, this variability is obvious.

### 2.1 Oakden's Analysis of Middle English Alliterative Poetry

J. P. Oakden published two books on Middle English alliterative poetry in 1930 and 1935, respectively, that were republished together as Oakden (1968). The first studied meter, and he gave counts (sometimes percentages) of the different types of lines for seventy-one poems, which includes all of the popularly studied examples in the literature. His data is used below, and although he is an expert, other specialists have performed counts that disagree with his. For example, Duggan reports (page 65 of Duggan (2000)) that his recount produced "results distinctly different from Oakden's." Consequently, the variability measurements below based on Oakden (1968) are only approximate.

Oakden uses the traditional coding where the forward slash stands for the caesura, letters stand for stressed words, and *x* means a non-alliterating stressed word as described in Section 1. Consequently, the first line of Figure 1 is "aa/aa," and the other nine are "aa/ax." Although there are only two patterns in the first ten lines, pages 186-7 of Oakden (1968) list 14 total meters for *Piers Plowman*, plus two additional categories: one for no alliteration at all and another called "complex groups," which includes infrequent, unusual patterns.

## 2.2 Alliteration Position Strings for Poetic Lines

The second coding in this paper is not used in literary circles but is a compromise between the more complex syllable scansion using / and *x* that is common in studies of prosody and the notation that only considers stressed words described in Section 2.1. Unfortunately, the author's attempts at automating the analysis of a poem by computer produced many errors, and doing it by hand had too many ambiguous cases, so a simplification was used. Each word with the initial alliterating sound, both stressed and unstressed, was coded with a 1, and all the other words were coded as 0. Using this, the lines of Figure 1 become 001101001, 0100100010, 01001000, 11001100, 00011010, 00010100, 011001000, 00111010, 00010101000, and 010010110. This is an approximation, however, and the following would be truer to the spirit of the alliterative revival, where 1s represent stressed alliterative words including those with unstressed prefixes: 001101001, 0100100010, 01001100, 01001100, 00011010, 01010100, 001101000, 00110010, 00010101000, and 010010100.

## 3. Defining the Mean and Variance of Categorical Data

Xavier Pennec worked out the core theory of doing statistical analyses with data fraom a Riemannian manifold, which is described in Pennec (1999) and Pennec (2006). He notes that the idea of generalizing the mean and variance is old and was published by Frechet in the 1940s (in French), but the extension to doing statistical testing seems to be new. His approach is two-fold. First, replace the Euclidean distance by the geodesic distance between two points. Second, define the mean and variance (and higher moments if desired) as an optimization problem. It turns out that this second step can be applied to string data using distance functions found in bioinformatics. The edit distance is used in this paper.

### 3.1 Pennec's Generalization of the Mean and Variance

The usual mean can be defined as the value, *c*, that minimizes *f(c)* as defined in Equation (3.1), and this minimum value is the variance.

$$f(c) = \frac{1}{n}\sum_{i=1}^{n}|x_i - c|^2 \qquad (3.1)$$

As Pennec (1999) points out, this approach of computing the mean and variance by solving an optimization problem has the advantage that it can be generalized to any Riemannian distance function, not just the usual Euclidean one. As Bilisoly (2013) notes, powers other than 2 can be used in Equation (3.1), which correspond to $L^p$ distances, and the most important case besides *p* = 2, is *p* = 1, which corresponds to the median. In this paper, only minimizing Equation (3.2) is done, where *d* is the edit distance, but note that any distance function could be used, and for each one there would be a corresponding generalized mean and variance.

$$f(c) = \frac{1}{n}\sum_{i=1}^{n}d(x_i, c)^2 \qquad (3.2)$$

### 3.2 Edit Distance

Edit distance between two strings was first analyzed in Levenshtein (1966) in the physics literature. The distance is defined to be the minimum number of edits to convert one string into the other, where insertion, deletion, and substitution of one character at a time are the only allowed operations. Section 3 of Bilisoly (2013) has the example of computing the distance between "old" and "halde" (two Middle English forms of the word "old" from McIntosh et al. (1986)). The distance is three because one can substitute *h* for *o*, insert an *a* and an *e*, but there is no way to do this in just two steps. This can be computed using dynamic programming in a number of steps equal to the product of the lengths of the strings. Although quadratic complexity is not fast enough for long strings (such as DNA sequences), this is not an issue in this paper.

Interestingly, the edit distance function, *d*, satisfies the axioms of a metric space. That is, for any three strings s1, s2, and s3, $d[s1, s2] \geq 0$; $d[s1, s2] = 0$ exactly when s1 = s2; $d[s1, s2] = d[s2, s1]$; and $d[s1, s3] \leq d[s1, s2] + d[s2, s3]$. There are many statistical ideas based on distance functions such as clustering, so this approach should prove fruitful.

### 3.3 A Simple Example

Before doing the more complex examples in Section 4, consider Figure 1 one more time. As noted above, those ten lines can be coded by the following strings: 001101001, 0100100010, 01001100, 01001100, 00011010, 01010100, 001101000, 00110010, 00010101000, and 010010100, where 1 stands for a stressed, alliterating word. These strings clearly vary because there is only one repetition, but now this can be quantified with a generalized variance.

The steps are as follows. First, the distance matrix in Figure 2 is computed, where each entry is the edit distance between two strings in the above list. Second, all the entries are squared, and then the rows are summed, which produces {94, 97, 75, 75, 66, 70, 70, 106, 104, 59}. Third, the minimum sum of squares is 59, the last entry, so the variance is 59/10 by Equation (3.2), and the mean string is the last one, 010010100.

One property of the above process is that the mean is one of the attested strings, which is unlike numerical data. For example, the average of a group of SAT scores is unlikely to be a multiple of ten, in which case it is not a score that could actually be obtained by a student. Although one could optimize over a larger space of strings, there is merit in having a mean string that codes a meter that actually has been used by the poet.

$$\begin{pmatrix} 0 & 4 & 4 & 4 & 3 & 3 & 1 & 3 & 3 & 3 \\ 4 & 0 & 3 & 3 & 3 & 3 & 4 & 3 & 4 & 2 \\ 4 & 3 & 0 & 0 & 2 & 2 & 3 & 4 & 4 & 1 \\ 4 & 3 & 0 & 0 & 2 & 2 & 3 & 4 & 4 & 1 \\ 3 & 3 & 2 & 2 & 0 & 3 & 3 & 2 & 3 & 3 \\ 3 & 3 & 2 & 2 & 3 & 0 & 3 & 4 & 3 & 1 \\ 1 & 4 & 3 & 3 & 3 & 3 & 0 & 2 & 2 & 3 \\ 3 & 3 & 4 & 4 & 2 & 4 & 2 & 0 & 4 & 4 \\ 3 & 4 & 4 & 4 & 3 & 3 & 2 & 4 & 0 & 3 \\ 3 & 2 & 1 & 1 & 3 & 1 & 3 & 4 & 3 & 0 \end{pmatrix}$$

**Figure 2:** Edit distance matrix for the first ten lines of *Piers Plowman*.

### 3.4 Generalizing the F Test

Once generalized variances are computed for two poems, these are almost never equal. However, this can be true because either they (1) have the same metrical structure or (2) use different meters. Statistical testing is used here to decide which of these is most likely to be correct.

One method for testing the equality of variances is the F test, the theory of which is well known. For example, Section 5.3 of Lehmann and Romano (2008) proves that Equation (3.3) gives the rejection region of the uniformly most powerful (UMP) unbiased test of $H_0: \sigma^2_y = \Delta_0 \sigma^2_x$ for bivariate normal data.

$$\frac{\Sigma(y_j - y_{mean})^2 / \Delta_0 (n_y - 1)}{\Sigma(x_i - x_{mean})^2 / (n_x - 1)} > c \quad (3.3)$$

Equation (3.3), however, is proportional to the ratio of the sample variances of the two variables. So the generalized version of this would be the ratio of the variances obtained by minimizing Equation (3.2). With this methodology, Section 4 compares *Sir Gawain* and *Piers Plowman* with respect to the codings in Sections 2.1 and 2.2.

## 4. Comparing *Sir Gawain* and *Piers Plowman*

The above ideas are straightforward to carry out in the software package, Mathematica. Because of (1) the difficulty of determining which words alliterate and (2) the size of the distance matrices, only the 220 non-Latin lines of the prologue of *Piers Plowman* are analyzed, and to make the samples equal in size, only the first 220 alliterating lines of Passus I of *Sir Gawain* are considered in Section 4.1. That is, the bobs and wheels are left out.

One simplification is to do computations only for the distinct coding strings, and then take into account the frequencies of these. However, this loses sequential information: that is, poetic lines have an order, and by computing the complete distance matrix, patterns across lines are possible to detect. Unfortunately, *Sir Gawain* has about 2000 lines and version B of *Piers Plowman* has about 7000 lines, which makes both of them too large to work with full distance matrices, so Section 4.2 uses the above simplification.

### 4.1 Analysis of Alliterating Position Strings

The coding in Section 2.2 (1 stands for any alliterating word, both stressed and unstressed, and 0 otherwise) is applied to the first 220 poetic lines, and an edit distance matrix is computed for both *Piers Plowman* and *Sir Gawain*. These are shown as color maps in Figures 3 and 4, where white denotes 0, and the darker the color, the larger the distance.

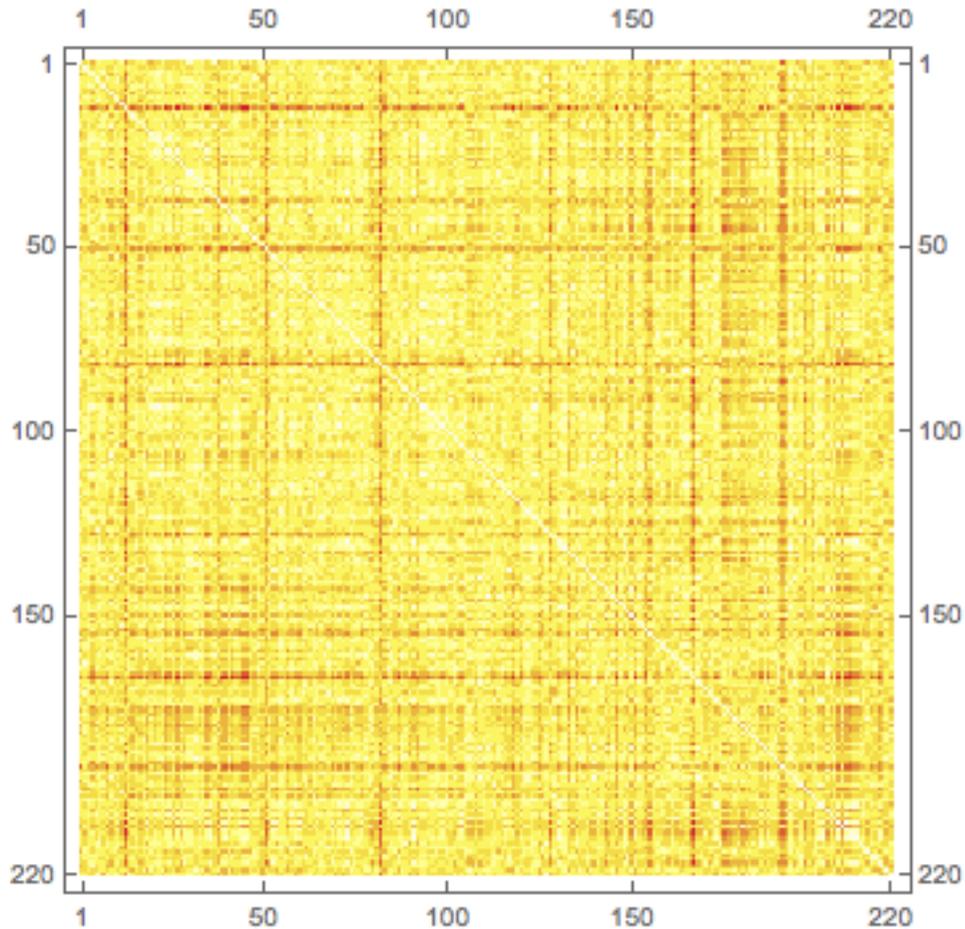

**Figure 3:** Map of the edit distance matrix of the prologue of *Piers Plowman* where the Latin lines have been removed. White stands for 0, and the darker the color, the larger the distance.

Both of these figures have dark rows (and columns) that indicate poetic lines dissimilar ton the rest. However, many of these are due to the coding system, which differs from the way an expert would scan a line. An extreme example of this is line 75 of *Sir Gawain*, which has an average value of 5.74 compared to the overall matrix average of 3.26. This anomaly arises because all of its words begin with vowels: "Of alderes, of armes, of oþer auenturus." Since all vowels alliterate and the difference between stressed and unstressed words was ignored, the code is "1111111," which is unlike the other binary vectors. Because of this situation, the results here are a proof-of-concept.

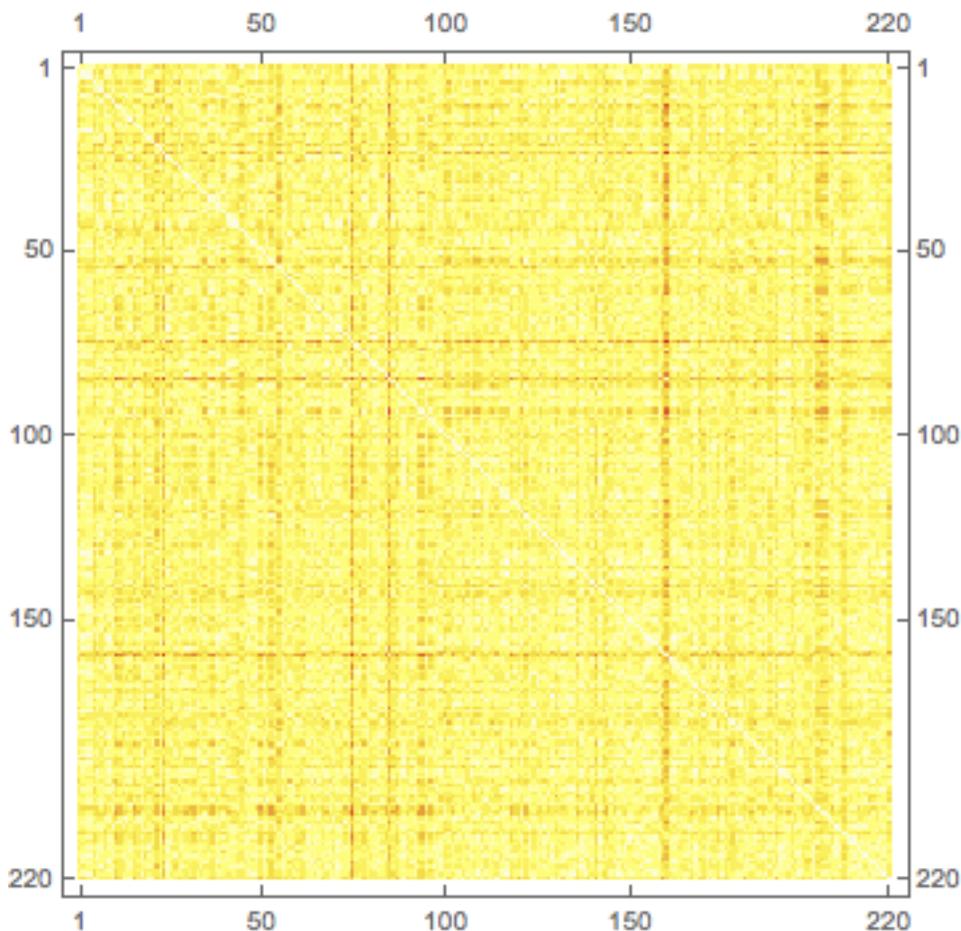

**Figure 4:** Map of the edit distance matrix of the first 220 lines of Passus I of *Sir Gawain*, where bob and wheels have been excluded. White stands for 0, and the darker the color, the larger the distance.

Given the edit distance matrices in Figures 3 and 4, the generalized means can be computed and are listed in Figure 5 (as the original poetic lines). Unlike the numerical mean, there need not be a unique answer. In addition, the generalized variances are 1601/220 for *Piers Plowman* and 1773/220 for *Sir Gawain*. Dividing the latter by the former gives 1773/1601 = 1.107, which is our test statistic. If this value is close enough to 1, then these two selections have the same prosodic structure, but otherwise they are significantly different. Resampling allows a p-value to be computed for this situation.

In statistical terms, one tests the null hypothesis, $H_0$: Both samples have the same prosodic variability. To do this, all the poetic lines are combined, randomly shuffled, and split into two subsets of 220 lines each. Edit distance matrices are recomputed for these two new "poems," and from these a new test statistic is produced. Finally, this shuffling process is repeated for a total of 1000 times, the results of which are given in Figure 6, and 1.107 is compared to this histogram. Since 94 of these 1000 values were either greater than 1773/1601 = 1.107 or less than 1601/1773 = 0.903, the empirical two-tailed p-value is 94/1000 = 0.094. This is not below $\alpha = 0.05$, so $H_0$ is not rejected. That is, based on this sample of poetic lines and the codings in Section 2.2 (which are less than one might desire), these two poets are plausibly using the same prosodic rules.

thanne loked up a lunatik a leene thyng withalle,
thanne greved hym a goliardeis a gloton of wordes,
ne carpynge of this coler that costed me nevere,
yet hoved ther an hundred in howves of selk there gode gawan watz grayþed gwenore bisyde,
for vch wyȝe may wel wit no wont þat þer were,
ther watz lokyng on lenþe þe lude to beholde

**Figure 5:** The generalized means for *Piers Plowman* (top 4) and *Sir Gawain* (bottom 3). The program has converted all the letters to lower case.

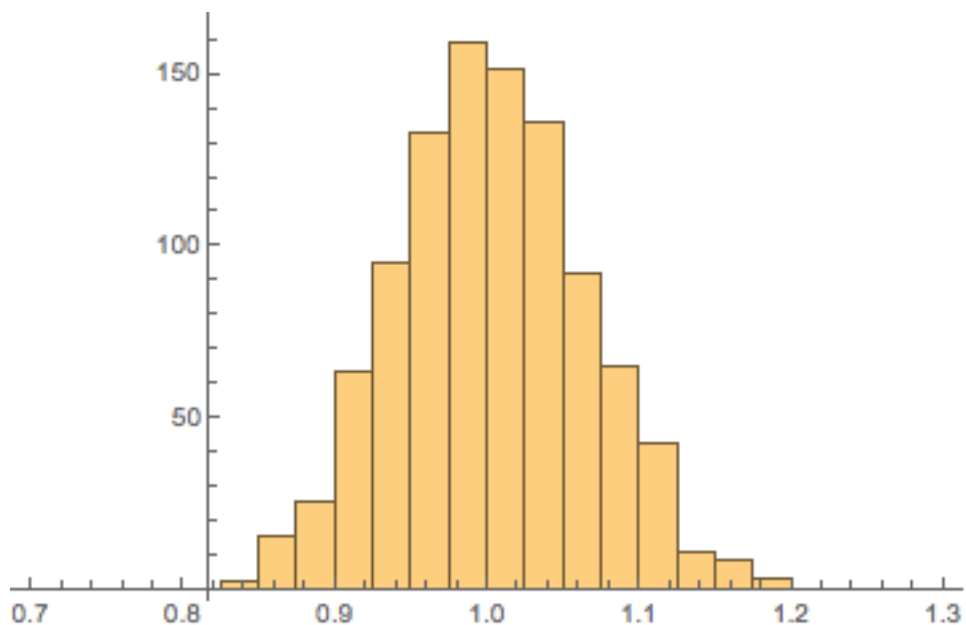

**Figure 6:** Empirical distribution of ratios of generalized variances for 1000 random shuffles of the 440 combined initial lines of *Piers Plowman* and *Sir Gawain* divided into two equal sized "poems."

**4.2 Analysis of Oakden's Alliteration Patterns**
Table 1 has the counts used in the analyses below. For *Sir Gawain* (SGGK), these came from pages 190-1. For *Piers Plowman* (PPB), these are given on pages 186-7 as percentages, which were converted to counts using the total number of lines, 7089, quoted on page 154. Because category 10, complex meters, only gave the combined percentage for *Piers Plowman*, these were dropped from both poems. Figure 7 shows the edit distance matrix, call it D, for the meters listed in Table 1. Put the counts for each poem in a diagonal matrix, say $C_{sggk}$ and $C_{ppb}$, and compute $DC_{sggk}$ and $DC_{ppb}$. Finally, repeat the analysis of Section 4.1 on these two products. That is, the row(s) with the minimum sum of squared entries is the generalized mean, and the value of this minimal sum divided by the number of poetic lines is the generalized variance.

| SGGK | Counts | PPB | Counts | Combined | Counts |
|---|---|---|---|---|---|
| aa/ax | 1532 | aa/ax | 4993 | aa/ax | 6525 |
| aa/xa | 23 | aa/xa | 117 | aa/xa | 140 |
| aa/aa | 70 | aa/aa | 538 | aa/aa | 608 |
| aaa/ax | 239 | aaa/ax | 291 | aaa/ax | 530 |
| aaa/xa | 4 | aaa/xa | 291 | aaa/xa | 295 |
| aaa/aa | 14 | aaa/aa | 64 | aaa/aa | 78 |
| ax/aa | 6 | ax/aa | 39 | ax/aa | 45 |
| xa/aa | 4 | xa/aa | 39 | xa/aa | 43 |
| ax/ax | 40 | ax/ax | 114 | ax/ax | 154 |
| xa/ax | 59 | xa/ax | 135 | xa/ax | 194 |
| aa/bb | 2 | aa/bb | 71 | aa/bb | 73 |
| ab/ab | 10 | ab/ab | 14 | ab/ab | 24 |
| ab/ba | 5 | ab/ba | 6 | ab/ba | 11 |
| aaa/xx | 2 | aaa/xx | 0 | aaa/xx | 2 |
| aa/xx | 0 | aa/xx | 241 | aa/xx | 241 |
| xx/xx | 0 | xx/xx | 50 | xx/xx | 50 |
| **SUM =** | **2010** | | **7003** | | **9013** |

**Table 1:** Counts given or inferred from Chapter 8 of Oakden (1968) for *Sir Gawain* (SGGK) and *Piers Plowman* (PPB).

$$\begin{pmatrix}
0 & 2 & 1 & 1 & 3 & 2 & 2 & 2 & 1 & 1 & 2 & 2 & 3 & 2 & 1 & 3 \\
2 & 0 & 1 & 3 & 1 & 2 & 2 & 2 & 3 & 3 & 2 & 3 & 2 & 2 & 1 & 3 \\
1 & 1 & 0 & 2 & 2 & 1 & 1 & 1 & 2 & 2 & 2 & 2 & 2 & 3 & 2 & 4 \\
1 & 3 & 2 & 0 & 2 & 1 & 3 & 3 & 2 & 2 & 3 & 3 & 4 & 1 & 2 & 4 \\
3 & 1 & 2 & 2 & 0 & 1 & 3 & 3 & 4 & 4 & 3 & 4 & 3 & 1 & 2 & 4 \\
2 & 2 & 1 & 1 & 1 & 0 & 2 & 2 & 3 & 3 & 3 & 3 & 3 & 2 & 3 & 5 \\
2 & 2 & 1 & 3 & 3 & 2 & 0 & 2 & 1 & 3 & 3 & 2 & 2 & 4 & 3 & 3 \\
2 & 2 & 1 & 3 & 3 & 2 & 2 & 0 & 3 & 1 & 3 & 3 & 3 & 4 & 3 & 3 \\
1 & 3 & 2 & 2 & 4 & 3 & 1 & 3 & 0 & 2 & 3 & 2 & 3 & 3 & 2 & 2 \\
1 & 3 & 2 & 2 & 4 & 3 & 3 & 1 & 2 & 0 & 3 & 3 & 4 & 3 & 2 & 2 \\
2 & 2 & 2 & 3 & 3 & 3 & 3 & 3 & 3 & 3 & 0 & 2 & 2 & 3 & 2 & 4 \\
2 & 3 & 2 & 3 & 4 & 3 & 2 & 3 & 2 & 3 & 2 & 0 & 2 & 4 & 3 & 4 \\
3 & 2 & 2 & 4 & 3 & 3 & 2 & 3 & 3 & 4 & 2 & 2 & 0 & 4 & 3 & 4 \\
2 & 2 & 3 & 1 & 1 & 2 & 4 & 4 & 3 & 3 & 3 & 4 & 4 & 0 & 1 & 3 \\
1 & 1 & 2 & 2 & 2 & 3 & 3 & 3 & 2 & 2 & 2 & 3 & 3 & 1 & 0 & 2 \\
3 & 3 & 4 & 4 & 4 & 5 & 3 & 3 & 2 & 2 & 4 & 4 & 4 & 3 & 2 & 0
\end{pmatrix}$$

**Figure 7:** The edit distance matrix for the combined meters in Table 1.

For *Sir Gawain*, the generalized mean is "aa/ax," which is not surprising because it is so common. The generalized variance is 71,011/2010 = 35.33. For *Piers Plowman*, the mean is also "aa/ax," and the variance is 1,352,636/7,003 = 193.15. Consequently, the test statistic, call it $F_{true}$, is 193.15/35.33 = 5.47. Finally, an approximate p-value is needed.

Instead of using the true meter counts in Table 1, the 2010 meters of *Sir Gawain* are combined with the 7003 meters of *Piers Plowman*, and shuffle these. The results are split into two pieces, one of size 2010, the other of size 7003. For each of these, a table of counts like the first four columns of Table 1 is created (note that the combined results will always be same.) Then the above calculation for $F_{true}$ is repeated for these new counts to produce $F_{resample}$. Figure 8 shows a histogram of 1000 $F_{resample}$s, and because all of these are smaller than $F_{true}$, the empirical one-sided p-value is 0/1000 = 0, which is obviously below α = 0.05. Hence the conclusion is that the two poems have different prosodic variabilities.

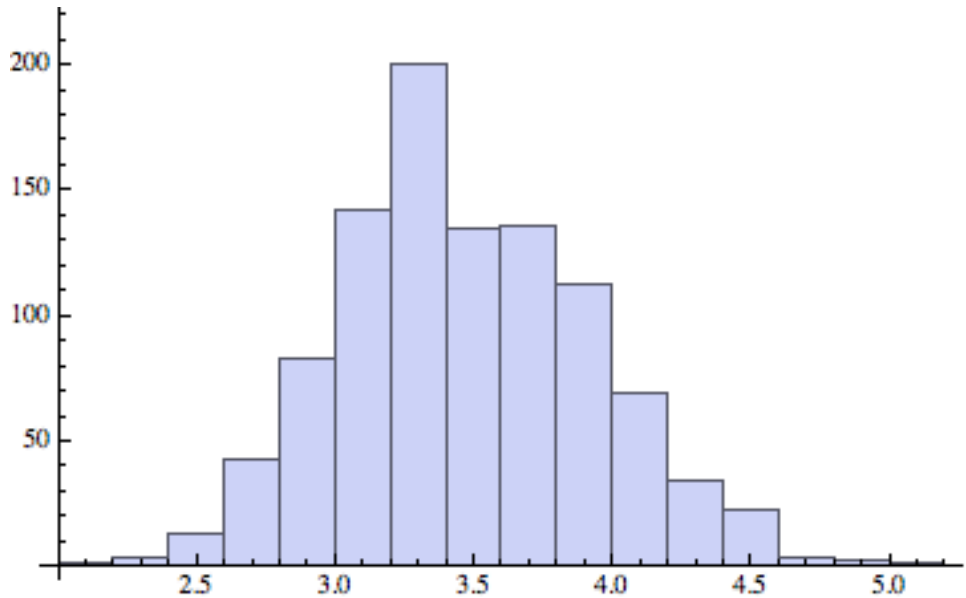

**Figure 8:** A histogram of 1000 $F_{resample}$s, the maximum of which is 5.07.

## 5. Discussion

Language data are tricky. For example, one might expect that Figure 8 would be centered about 1, not a value near 3.5. This is partly because the two poems have much different lengths and is influenced by the highly skewed distribution of meters in Table 1. For example, for word counts, it is well known that dividing by the sample size to create proportions is not enough to remove the effects of sample size, which happens because of the skewness of the counts. This is discussed in detail in Baayen (2002).

Nonetheless, this paper shows that the above methodology gives answers even with large poems such as *Sir Gawain* and *Piers Plowman*. Moreover, it can be used with any distance function, so it is a flexible approach.